\documentclass{llncs}
\usepackage{makeidx}  
\usepackage{subfigure}
\usepackage{multirow}
\usepackage{epsfig} 
\usepackage{color}
\usepackage{booktabs}
\usepackage{amsmath} 
\usepackage{amssymb}  
\usepackage{epstopdf}
\begin{document}

\title{Towards Data-Driven Hierarchical Surgical Skill Analysis}
\author{Bin Li \inst{1} \and B\`{e}r\`{e}nice Mettler \inst{1} \and Timothy M. Kowalewski \inst{2}}
\institute{ (1) Department of Aerospace Engineering and Mechanics and (2) Department of\\Mechanical Engineering, University of Minnesota, Minneapolis, MN, 55455, USA\\
{\small \{lixx1778, mettler, timk\}.umn.edu} }

\maketitle

\begin{abstract}
This paper evaluates methods of hierarchical skill analysis developed in aerospace to the problem of surgical skill assessment and modeling. The analysis employs tool motion data of Fundamental of Laparoscopic Skills (FLS) tasks collected from clinicians of various skill levels at three different clinical teaching hospitals in the United States.
Outcomes are evaluated based on their ability to provide relevant information about the underlying processes across the entire system hierarchy including control, guidance and planning.
\end{abstract}
\section{Introduction}

Over 32,000 deaths and \$9B in losses are annually attributed to avoidable surgical errors~\cite{zhan2003excess},  highlighting the need to ensure quality by mandating proficiency benchmarks in standardized training and credentialing~\cite{sturm2008systematic} of surgical trainees. Modern procedural trainees and graduates prove inadequately prepared and in need of additional training~\cite{mattar2013generalsurgeryresidency,Coleman2013early}. Yet this is not viable with the faculty-intensive and time-demanding, subjective methods in use today, particularly given the steady influx of ever-changing technologies into the operating room.  Automated, objective assessments of skill are needed. To ultimately address such a broad issue, simple performance scores on tasks such as task time or path length will not suffice,  the entire, complex spectrum of human skill must be treated.

\subsection{Surgical skill analysis}
Prior work~\cite{reiley2010review,varadarajan2009data} has made considerable advances in task segmentation and skill classification for surgical contexts, particularly~\cite{kumar2012assessingsystemoperation,zappella2013surgicalgestureclassification,ahmidi2013stringmotifbased,tao2012sparsehiddenmarkov,tao2013segmentationandrecognition}. These approaches tend to focus on a specific subtask or modality (e.g., robotics vs. manual laparoscopy).
While they succeed in providing valuable metrics to discriminate skill or procedural context, they do not extend directly to hierarchical human skill constructs like perception, planning, and cognition that are ultimately vital to this area~\cite{gallagher2005virtual}. No comprehensive framework exists that can successfully tie these many disparate attributes.  Universal metrics of skill, proposed in~\cite{kowalewski2012realtime}, provided an early approach to such task-agnostic metrics  and produced datasets to ultimately evaluate such metrics, however, it yielded little progress towards such a goal.  We herein introduce a different approach based on invariants. The approach makes it possible to delineate between key processes of the  hierarchical control and sensory system. We provide preliminary evaluation for laparoscopy.

\subsection{Alternative Skill Model and Analysis Framework} \label{sec:framework}

More recently researchers have grown interested in a more formal dynamics and control based theory of perception and action. The notable examples include Warren's control theory of dynamics of action and perception~\cite{Warren04,warren2006dynamicsofperceptionaction}. 
More comprehensive models that capture the closed-loop interaction have been proposed in the aerospace field in the form of multi-loop models. The loops are organized hierarchically starting with the low-level attitude stabilization, to tracking, and ultimately goal directed maneuvering~\cite{Heffley1979manualappraochtohover,Hess1986automationeffectsmultiloopmanual,White2004adaptivepilotmodel}.
These models and efforts suggest that comprehensive skill evaluation requires accessing and using information across the different levels of the system hierarchy; not just the performance ``outputs'' but also the various internal processes and if possible should encompass the ``inputs'' to the system such as perception and attention.

\begin{figure}[h!]
  \centering
  \includegraphics[width=.5\columnwidth]{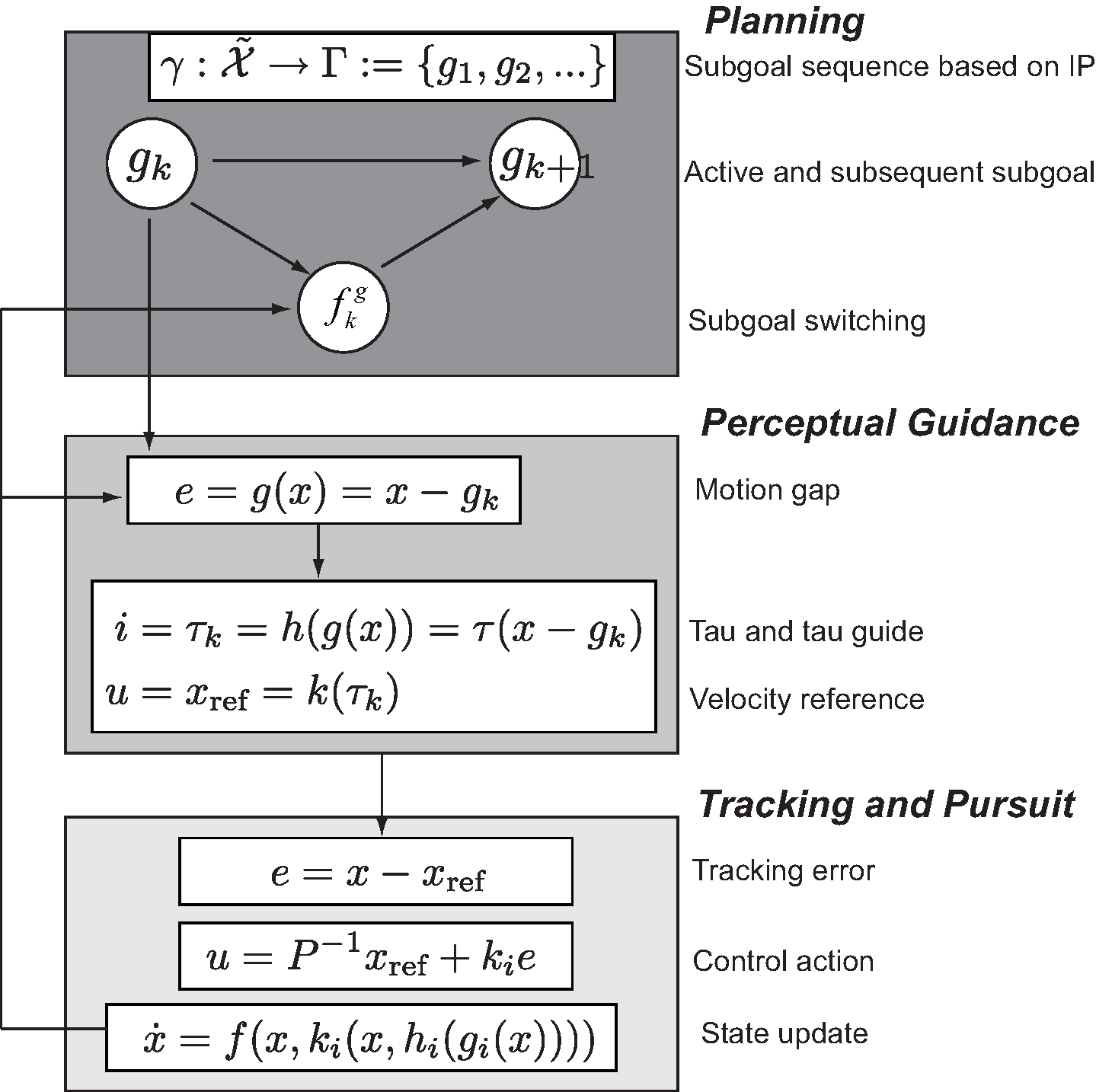}
  \caption{Hierarchic multi-loop model of human guidance behavior. The top level describes the planning level based on the decomposition of the task and environment in terms of interaction patterns. The plan is codified based on a subgoal sequence $g_k$. The currently active subgoal defines the reference for the perceptual guidance. The latter extracts the current motion gap which is used to determine a state reference trajectory $x_\text{ref}$. At the lowest level, a tracking feedback system implements the desired motion. }
  \label{fig:hierarchical_model}
\end{figure}

The multi-loop framework provides a rigorous, deterministic basis for measurement, evaluation and modeling of skills. Figure \ref{fig:hierarchical_model} shows the primary loops in a multi-loop model~\cite{MettlerKong12}. This hierarchical multi-loop model suggests that operators learn feedback structures across multiple levels.

The concept of ``interaction pattern'' is introduced in~\cite{KongMettler13agentenvironmentinteractions} following investigation of human guidance behavior using experiments with miniature remote control helicopters. These patterns are based on invariants of the closed-loop interactions. The significance of these invariants is that they describe what principles human pilots or operators use to break down complex guidance problems into a sequence of smaller, tractable ones.
Preliminary studies based on piecewise affine (PWA) model identification methods suggest that the equivalence classes can be further decomposed into distinct dynamic modes, which provides deeper insight into lower-level control strategies.
The higher-level interaction patterns combined with the lower-level dynamic modes provide the building blocks needed to codify the behavior across the entire hierarchy, from the lower-level control, guidance and perception, all the way to higher-level planning, adaptation and learning~\cite{MettlerKong12}. This paper investigates how this framework can be used for surgical skill evaluation.

\section{Experimental Setup} \label{sec:exp}
\begin{figure}[h!]
  \centering
  \subfigure[EDGE]{
    \includegraphics[width=0.3\columnwidth]{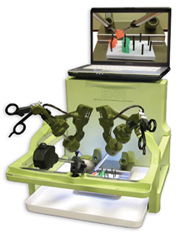}
    \label{Fig:EDGE}
  }
  \subfigure[Peg Transfer Task]{
    \includegraphics[width=0.5\columnwidth]{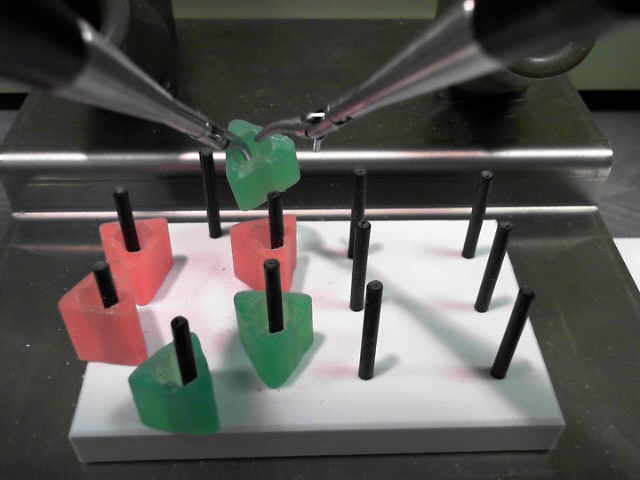}
    \label{Fig:PegTransferScreenshot}
  }
  \vspace{-0.6em}
  \caption{The EDGE platform (a) and screen shot of the FLS Peg Transfer task (b) used for this work.}
  \label{Fig:EDGEandPegTx}
\end{figure}

This study employed the dataset collected in~\cite{Kowalewski2012} which used the Electronic Data Generation and Evaluation (EDGE) platform (Simulab Corp. Seattle, WA), Fig. \ref{Fig:EDGE}. This consists of 22.7 hours of synchronized video and tool motion data of Fundamental of Laparoscopic Skills (FLS) tasks collected from clinicians of various skill levels at three different clinical teaching hospitals in the United States. FLS  has been shown to correlate to operating room performance~\cite{sroka2010fundamentals}.   We herein incorporate only a small part of this data that also provides ratings by faculty clinicians via blinded video review to establish valid categories of skill.

\subsection{Task Description}
In this paper, we use only the Peg Transfer task 
where clinicians use Maryland Graspers to transfer blocks in minimal time and with minimal drops.  The blocks must be picked up by one hand with a laparoscopic tool and then transferred mid-air to the other hand tool.

\subsection{Data Overview and Group Selection}
The motion data (tool tip position, orientation, grasp angle and grasp force for both hands) sampled at 30Hz.  Three skill groups were selected based on a combination of criteria in Table \ref{table:GroupSelectionOverview}. Complete details are available in~\cite{kowalewski2012realtime}.

A set of six complete Peg Transfer task instances were arbitrarily selected from unique subjects among the three geographically distinct  sites to represent each of the three skill groups.
\begin{table}[htbp]\small
\caption[Group Selection Overview]{ Summary of criteria used to select each set of iterations and its intended purpose. $N$ refers to the total count of iterations of each set.  }\label{table:GroupSelectionOverview}
\centering
\begin{tabular}{ccp{2in}}
\toprule
\bfseries Group &\bfseries $N$ &\bfseries Criteria  \\ 
\midrule
Expert\newline(Exp) 	& $6$ & {\small{Practicing laparoscopists (over 100 lapr. procedures): surgeons' and fellows' best FLS-scoring logs with 3/5 or greater average OSATS video review scores.  }}  \\[8pt]
Intermediate (Int) 	& $6$ & \small{15th percentile  of FLS scores about midpoint FLS score determined between lowest Expert FLS score and highest Novice FLS score $[.59, .73]$.}\\[8pt]
Novice (Nov) 		& $6$ & \small{All logs below $15^{th}$-percentile FLS score. }     \\
\bottomrule
\end{tabular}
\end{table}

\section{Surgical performance analysis} \label{sec:analysis}

\subsection{Task-Level Statistical Analysis}
\label{sec:TaskStats}
\begin{figure}
  \centering
  \includegraphics[width=0.8\textwidth]{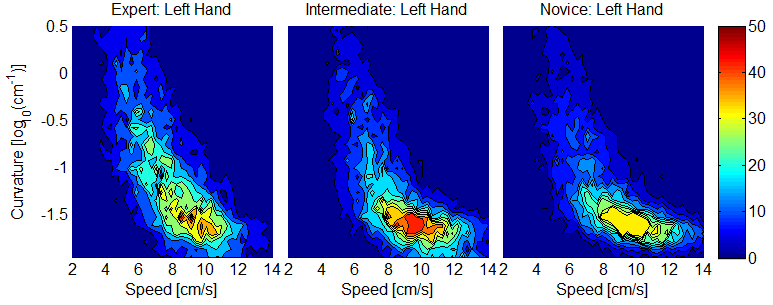}\\
  \vspace{-0.3cm}
  \caption{2D distribution of speed and curvature}\label{Fig:pdf}
\end{figure}

The objective of statistical analysis at the task level is to provide general characteristics of operator skill in an intuitive context. A mapping of the surgical movements into a probability distribution in the speed-curvature space was adopted as initial technique to assess general characteristics of operator\rq{}s performance. This analysis provides both the maneuver envelope and the dominant states in the behavior~\cite{Li2013}. The dominant states are defined to be the most frequently visited states which may serve as the transition quasi-equilibrium between maneuvers.
The probability distribution of trajectory points in the speed-curvature plane is shown in Figure~\ref{Fig:pdf}. The expert group exhibits a larger envelope and more condensed dominant states.

Probability distributions generated over the same event space can be compared via a symmetric Kullback-Leibler divergence $D(p||q) = [D_{KL}(p||q) + D_{KL}(q||p)]/2$.
A leave-one-(surgeon)-out permutation analysis was performed for all three skill groups.  The resulting divergence among the populations of possible group probability distributions is shown in Table \ref{table:GrpPermutation}. Each entry in the table, $e_{m,n}$, is the mean and standard deviation of
${D(P_{grp_m \backslash subj_i }||P_{grp_n \backslash subj_j })}, \forall i \ne j$. Low values along the diagonal indicate that group distributions remain consistent and remain distinct from other groups even if individual surgeons are removed, whereas the low standard deviations (in parenthesis) indicate little change in overall distribution due to the removal of a single surgeon.  However,  the classification power of such a distributional approach is poor, see Table \ref{table:ConfusionMtx}. This shows how such broad statistical approaches succeed in providing general characteristics of operator behavior in an intuitive context (the task variables), however, such generalization is too broad to be used for classification of very specific runs or individuals.

\begin{table}[htbp] 
\caption{Leave-one-(surgeon)-out (a) permutation analysis for symmetric KL-Divergences of group distributions and (b) cross-validation confusion matrix for classification success for individual surgeons. }
\label{table:Confusion}

\centering
\subtable[Permutation Analysis, mean (std dev) ]{
\label{table:GrpPermutation}
\begin{tabular}{cccc}
\toprule
\parbox[t]{1.5cm}{ \quad }& & Alternate &\\
\footnotesize{ } & \bfseries Exp &\bfseries Int &\bfseries Nov  \\ 
\midrule
\bfseries Exp  &0.18 (0.10)   &	0.82  (0.03) &	0.78  (0.05) 	 \\
\bfseries Int  	&0.81 (0.02) &	0.14  (0.04) &	0.81  (0.01) 	 \\
\bfseries Nov  &0.86 (0.05) &	0.69  (0.05) &	0.11  (0.02) 	 \\
\bottomrule
\end{tabular}
}
\qquad
\subtable[Classifier Performance]{
\label{table:ConfusionMtx}
\begin{tabular}{ccccc}
\toprule
&& Predicted  &\\
\footnotesize{ } & \bfseries Exp &\bfseries Int &\bfseries Nov  \\ 
\midrule
\bfseries Exp  & 3 &  2 &  1 & \\
\bfseries Int  & 0 &  4 &  2 & \\
\bfseries Nov  & 1 &  2 &  3 & \\
\bottomrule
\end{tabular}
}%

\end{table}

\vspace{-5mm}

\subsection{Kinematic classification}
\label{sec:KinematicClassification}
\begin{figure}[ht]
  \begin{minipage}{.495\textwidth}
    \centering
    \subfigure[Trajectory]{
      \includegraphics[width=0.45\columnwidth]{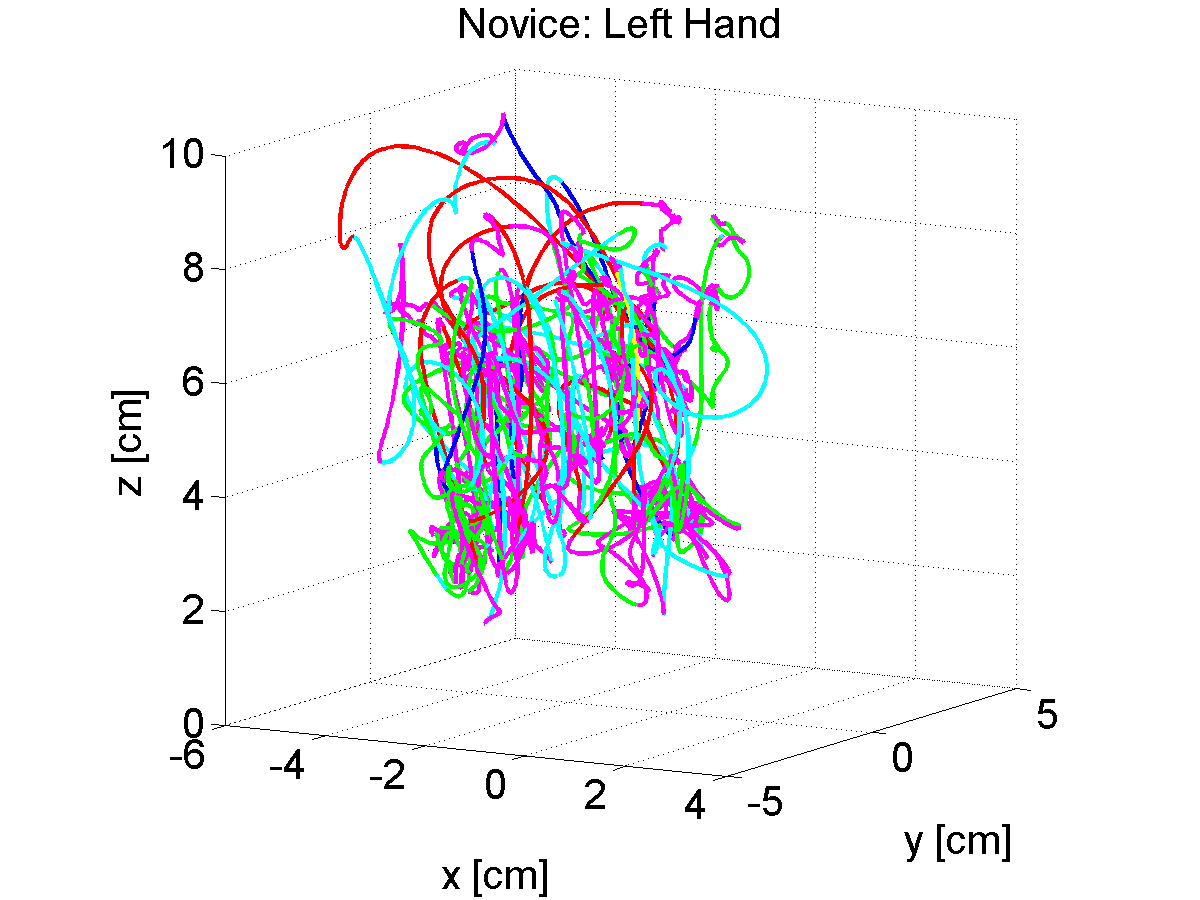}
      \label{Fig:mp_traj_novice}
    }
    \subfigure[Ratio]{
      \includegraphics[width=0.45\columnwidth]{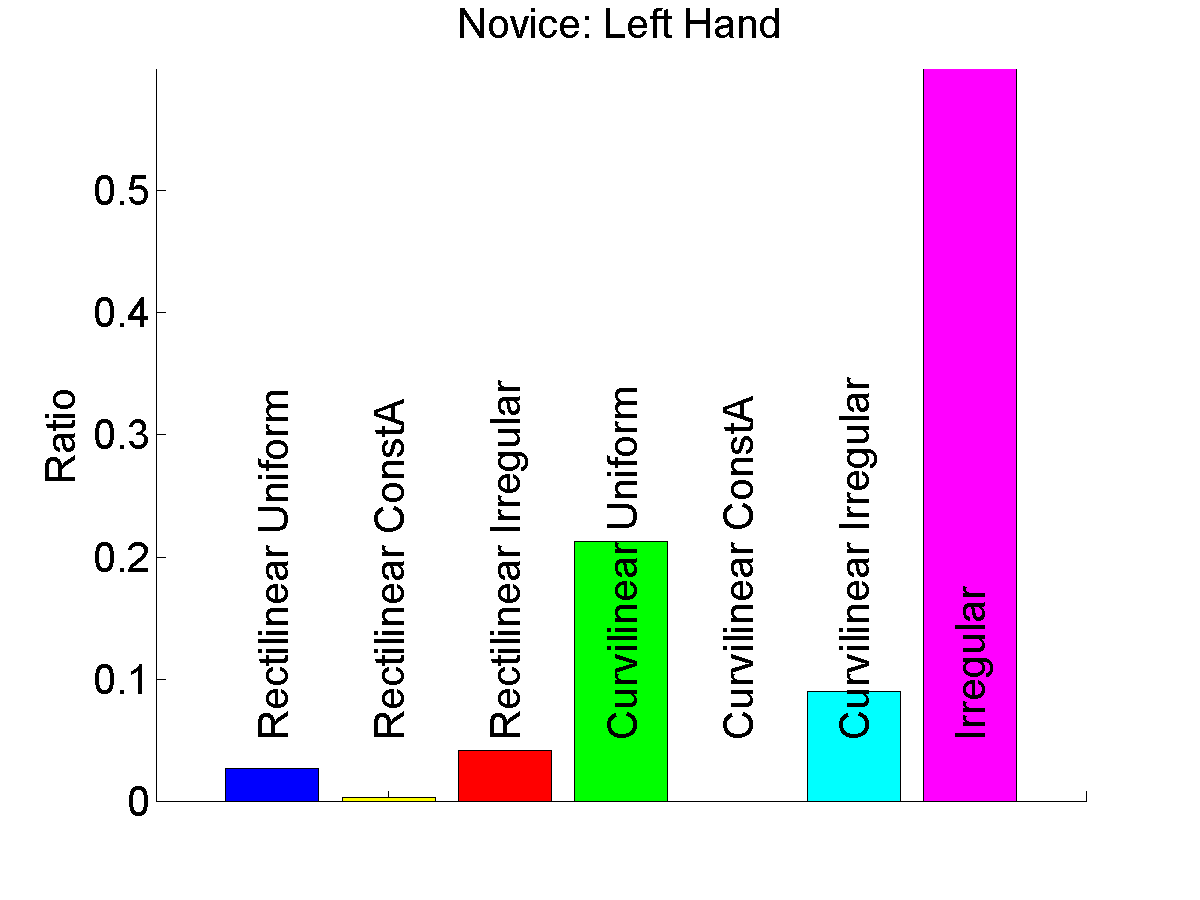}
      \label{Fig:mp_ratio_novice}
    }
    \vspace{-0.6em}
    \caption{Novice kinematic classification}
    \label{Fig:mp_novice}
  \end{minipage}
  \begin{minipage}{.495\textwidth}
    \centering
    \subfigure[Trajectory]{
      \includegraphics[width=0.45\columnwidth]{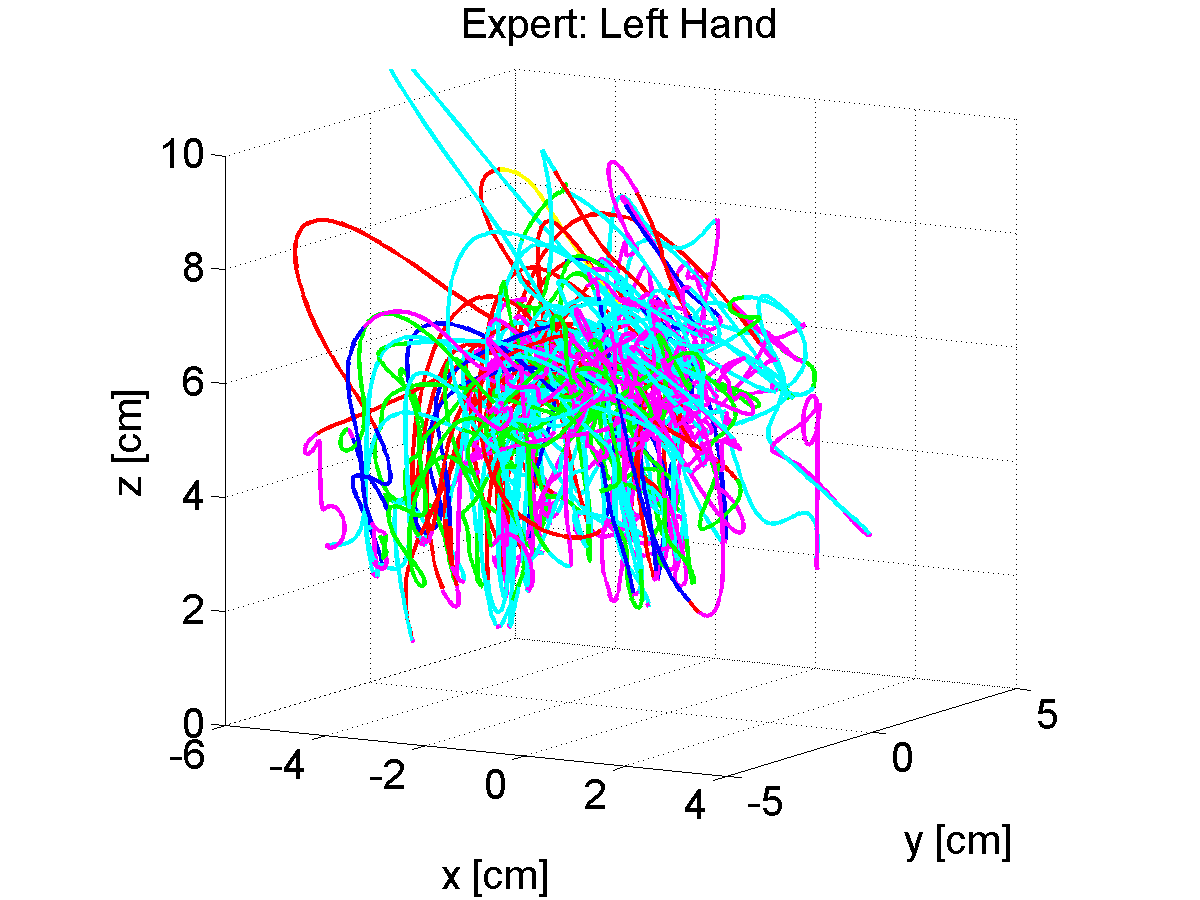}
      \label{Fig:mp_traj_expert}
    }
    \subfigure[Ratio]{
      \includegraphics[width=0.45\columnwidth]{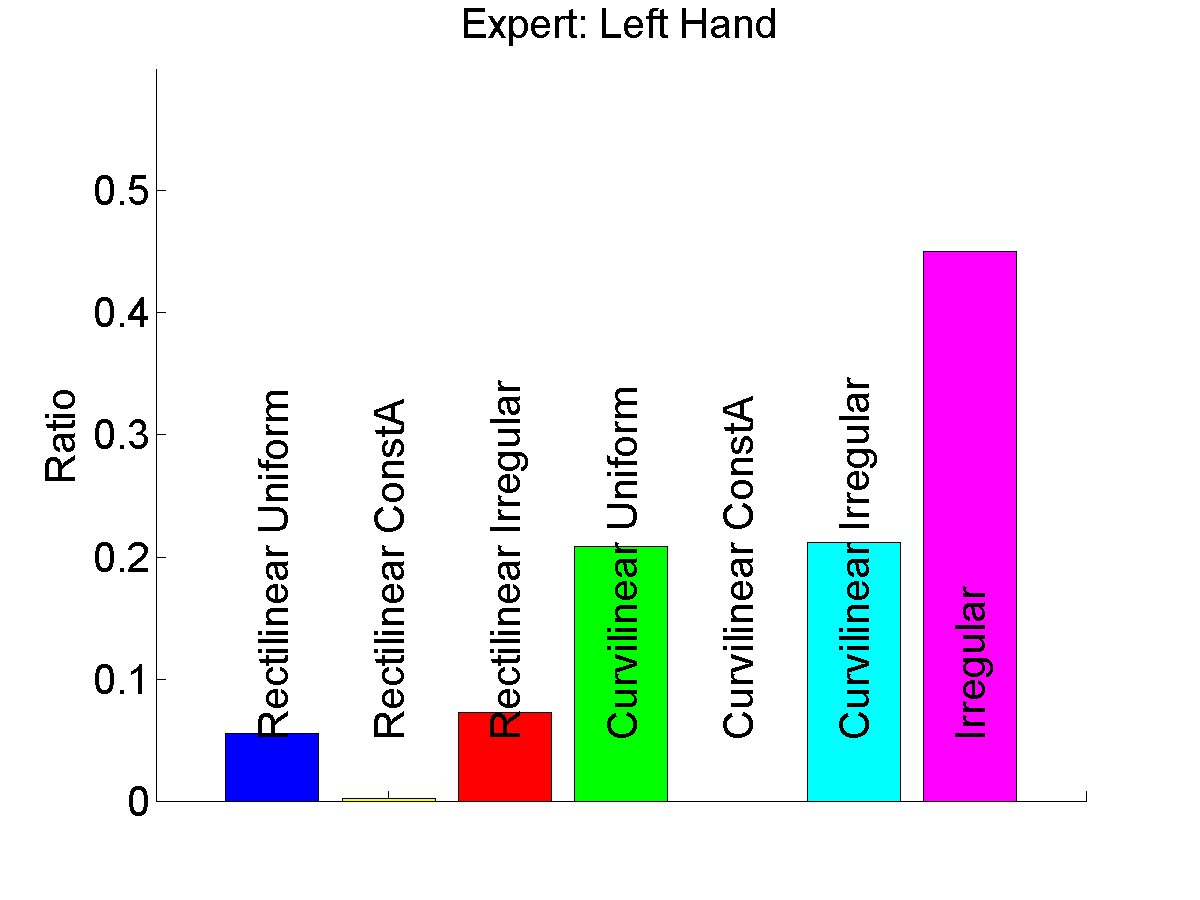}
      \label{Fig:mp_ratio_expert}
    }
    \vspace{-0.6em}
    \caption{Expert Kinematic classification}
    \label{Fig:mp_expert}
    \end{minipage}
\end{figure}

The kinematic classification method introduced in~\cite{Li2013} is based on the concept of control and attention workload. The method uses a library of motion primitives of different attention load levels.
The assumption is that experts favor motions of low control and attention load. Simpler motions, such as rectilinear and uniform (non-accelerated) motion, are easier to implement and more predictable, and therefore demanding less attention. These motions would allow the maneuvers to be more efficient and more consistent in multi-trial operations. Moreover, given the limited information processing capacity of human, simpler motions would allow for more cognitive processes, in particular, planning and decision making.
Parsing trajectories under human control into sub-level sequences of motion primitives provides insights into the organization principles which is an essential aspect of human spatial control skills.
The metrics derived from the segmentation, include the frequency of motion primitives and the mean duration of each segments.
The kinematic classification results are shown in Figures~\ref{Fig:mp_novice} and~\ref{Fig:mp_expert}.

\subsection{Dynamic clustering}

The dynamic clustering method is based on the assumption of hierarchic organization of behavior~\cite{Li2013}. It assumes that humans tend to adopt limited number of strategies in dealing with complex interactions taking place between organism, task and environment elements. The interaction involves the whole system of processes including perception, cognition and motor control. 
With extensive practice, these interactions will exhibit patterns that are manifestation of processes used to reduce the attention load and facilitate the organization of behaviors. Therefore capturing and describing these patterns are significant to investigate skills across the comprehensive hierarchy of processes.

\begin{figure}[th]
  \centering
  \begin{minipage}{.495\textwidth}
    \centering
    \subfigure[Trajectory]{
      \includegraphics[width=0.45\columnwidth]{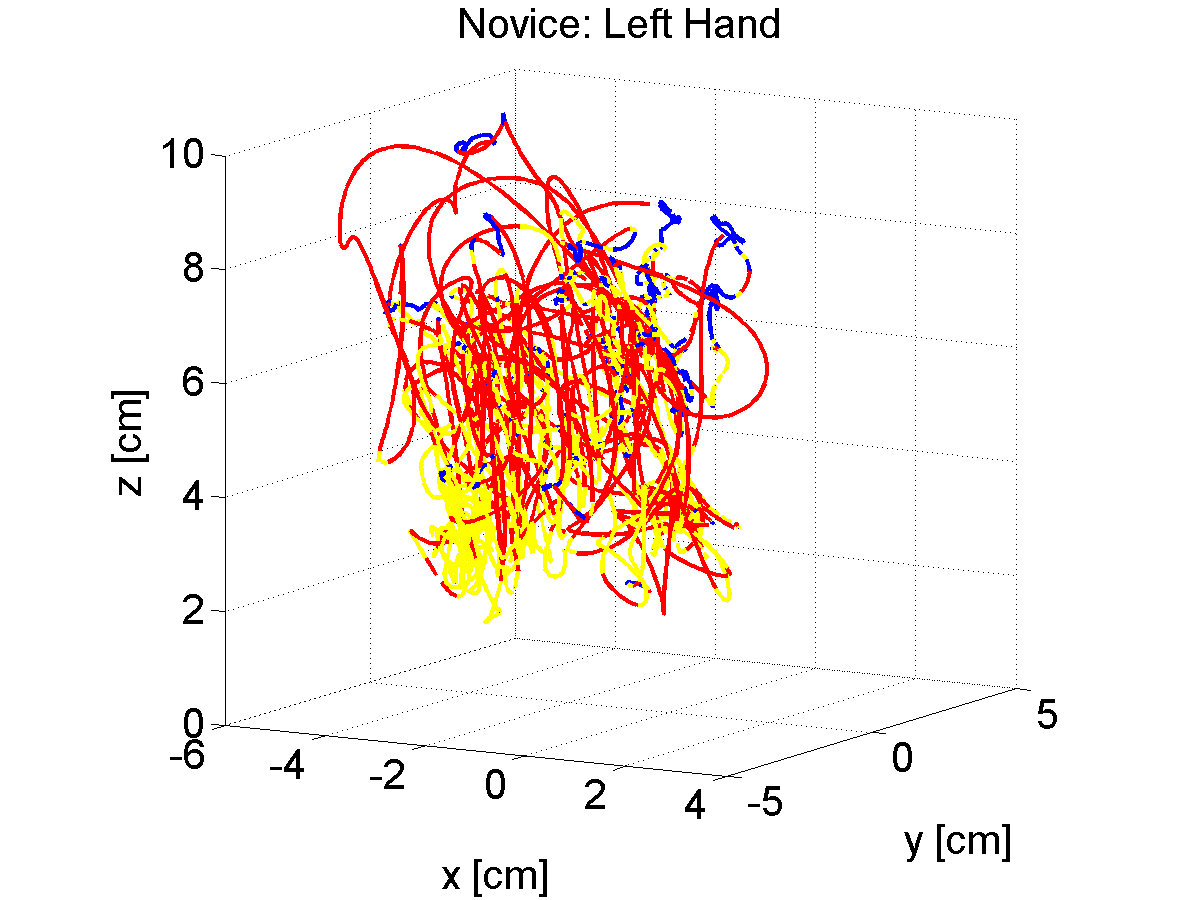}
      \label{Fig:pwa_traj_novice}
    }
    \subfigure[Ratio]{
      \includegraphics[width=0.45\columnwidth]{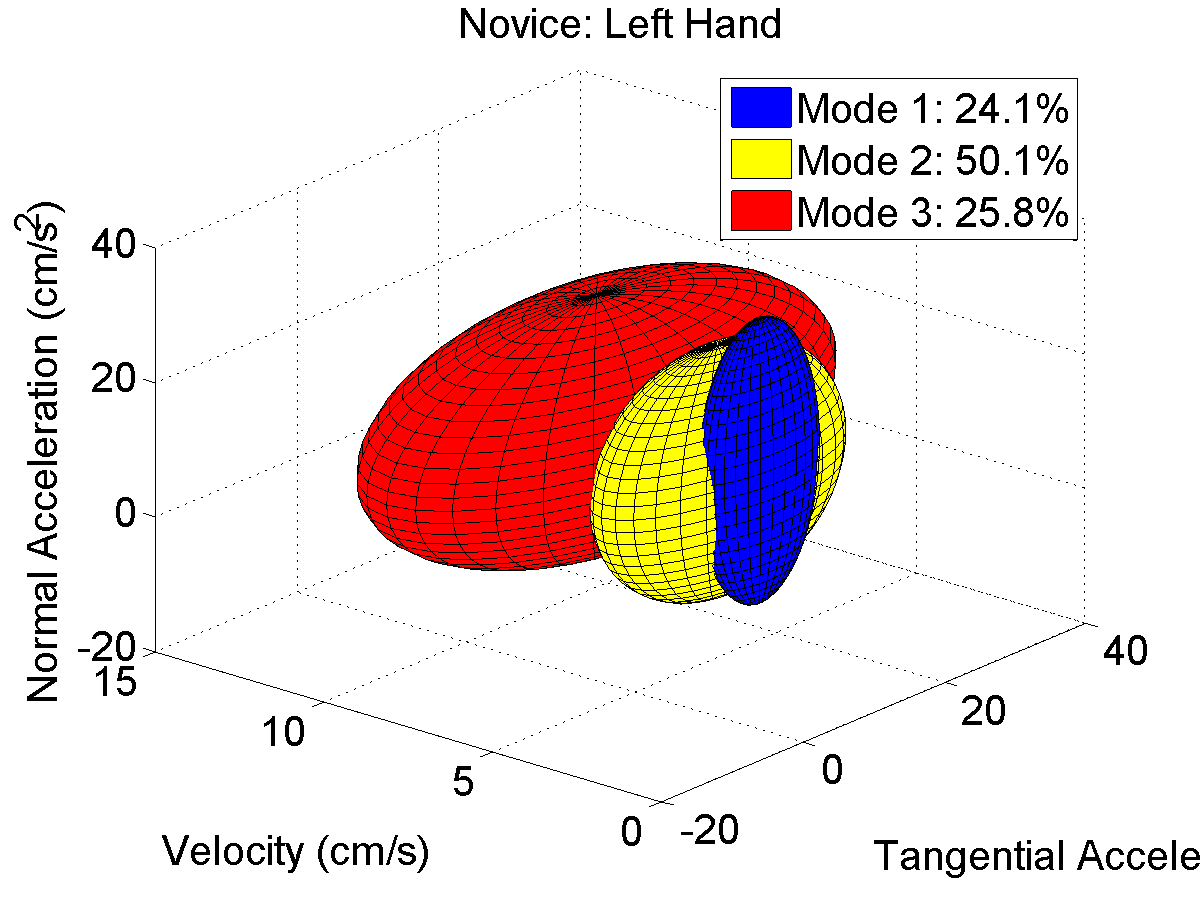}
      \label{Fig:pwa_ratio_novice}
    }
    \vspace{-0.6em}
    \caption{Dynamic clustering of novices}
    \label{Fig:pwa_novice}
  \end{minipage}
  \begin{minipage}{.495\textwidth}
    \centering
    \subfigure[Trajectory]{
      \includegraphics[width=0.45\columnwidth]{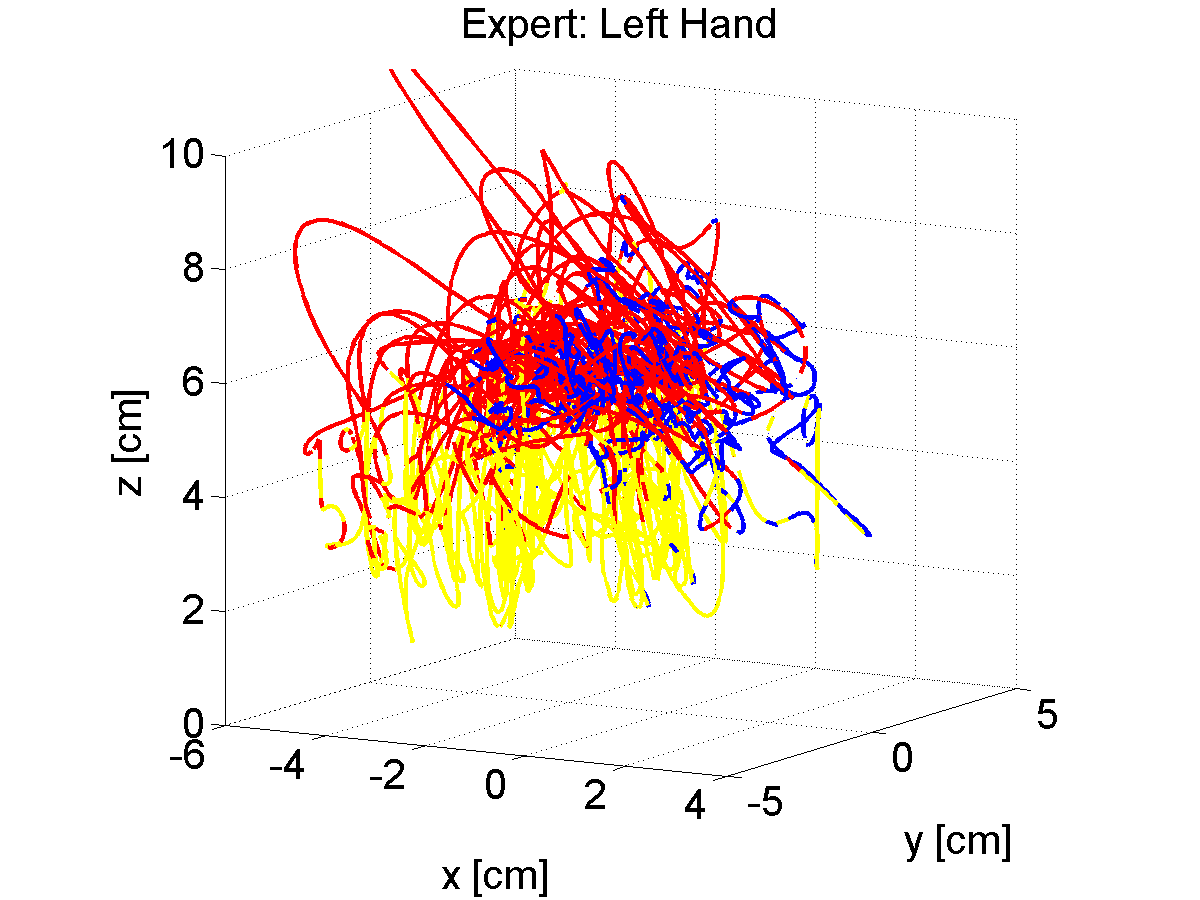}
      \label{Fig:pwa_traj_expert}
    }
    \subfigure[Ratio]{
      \includegraphics[width=0.45\columnwidth]{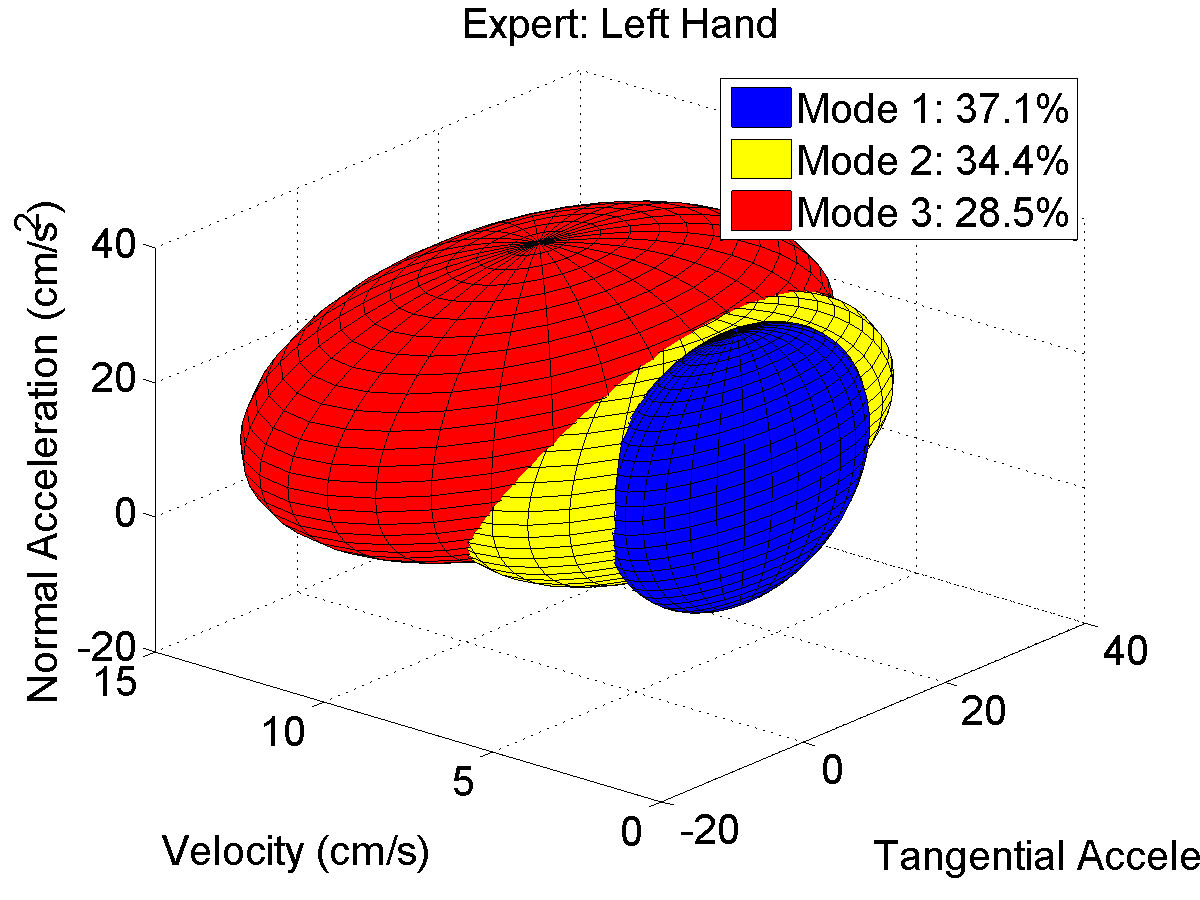}
      \label{Fig:pwa_ratio_expert}
    }
    \vspace{-0.6em}
    \caption{Dynamic clustering of experts}
    \label{Fig:pwa_expert}
  \end{minipage}
\end{figure}

In the dynamic clustering method, the dynamics of human-agent-environment interaction is described with Piece-Wise Auto-Regressive eXogenous (PWARX) model given in a parametric state-space form. Although the closed-loop interaction dynamics is always non-linear, it assumes that each interaction pattern describes an invariant in human's behavior that manifests as a quasi-equilibrium in the dynamics. Therefore, the interaction patterns can be captured using a PWARX model in the form described in~\cite{Li2013} and identified with different set of parameters:
\begin{eqnarray}\label{Eq:M2}
    \begin{bmatrix}
    x\\
    y\\
    v_x\\
    v_y
    \end{bmatrix}_{k+1}
    =
    \begin{bmatrix}
    1 & 0 & \Delta t & 0\\
    0 & 1 & 0 & \Delta t\\
    a_{31} & 0 & a_{33} & 0\\
    0 & a_{42} & 0 & a_{44}
    \end{bmatrix}
    \begin{bmatrix}
    x\\
    y\\
    v_x\\
    v_y
    \end{bmatrix}_k
    +
    \begin{bmatrix}
    0\\
    0\\
    b_3\\
    b_4
    \end{bmatrix}
\end{eqnarray}
where $\Delta t$ is the sampling time~\cite{Li2013}. Each interaction pattern is identified as a model of different set of parameters.
The PWARX results are shown in Figures~\ref{Fig:pwa_novice} and~\ref{Fig:pwa_expert}, where three clusters are identified for both novice and expert groups. To make the analysis more intuitive, the PWARX parameters are transformed into speed, normal and tangential acceleration in the ellipsoid plot.

\section{Hierarchical Skill Assessment Results}

\subsection{Subgoal closure}
As suggested in the introduction, dynamics-based clustering reveals spatial organization abilities of expert surgeons in performing Peg Transfer task, and makes it possible to delineate the different phases of the task and therein analyze specific performance characteristics. In Figure~\ref{Fig:pwa_traj_expert}, the spatial organization of the behavior of expert surgeons is closely correlated with the three phases in Peg Transfer task:
\begin{enumerate}
  \item Starting phase (cluster mode 2) coincides with the surgeons picking up the blocks. The movement during this phase follows a medium velocity range.
  \item Maneuvering phase (cluster mode 1) coincides with the surgeons moving the gripped blocks to the central area of the board. There is no restriction on the movement during the maneuvering phase, and the objective of the phase is to be as fast as possible. Therefore the surgeons adopt high velocity and the accelerations span a large range.
  \item Interception phase (cluster mode 3) coincides with the blocks being transferred in the air between two hands of laparoscopic tool. This phase is critical in that it
      requires a large amount of coordination effort for both hands.
\end{enumerate}

For each phase of the task, expert surgeons adopt very consistent strategy. In contrast, the maneuvers of novices are less consistent. During the starting phase, novices sometimes drive the control to high velocity, which penalizes the accuracy. The lack of consistency in the strategy also demands more attention load to plan for new trials and to handle the range of conditions. More attempts are required for novices to successfully pick up the blocks. In the maneuvering phase, the laparoscopic tool frequently slows to a lower velocity, penalizing the completion speed of the task.

\subsection{High-level planning}

To facilitate the accomplishment of complex task, humans divide the task into subtasks. In~\cite{KongMettler13agentenvironmentinteractions}, Kong and Mettler have shown that subtasks exploit invariants in the dynamic environment interactions. The invariants in the human behavior emerge through extensive practice ostensibly as a result of the assimilation of coordinated movement and perceptual processes in procedural memory. These interaction pattern can then be used as a unit of behavior for the larger organization. High-level planning can therefore be assessed from the organization of interaction patterns. Effective planning allows using interaction patterns that take advantage of the dynamic interaction between human\rq{}s motor skills and task elements that also reduce the attention load. For this reason, spatial organization of the interaction pattern is an important measure of the surgeons' planning skill. To quantify the spatial organization, the Cartesian coordinates of trajectory points are classified using a Fisher classifier based on the tags obtained in PWA clustering.
The misclassification ratio is then used as the measure of spatial organization, as shown in Table~\ref{Tab:spatial_org}.

\begin{table}
  \centering
  \caption{Skill metric in high-level planning}\label{Tab:spatial_org}
  \begin{tabular}{c|ccc}
    \hline
     Spatial organization [\%] & Expert & Intermediate & Novice \\ \hline
      Complete Groups & 17.9 & 29.6 & 38.6\\
      Leave-One-Out Mean(std dev) & 13.3  (4.4)  & 27.1 (5.6)  &  35.0 (6.5) \\
  \end{tabular}
\end{table}

\section{Conclusion} \label{sec:conclusion}
The results underscore the limitations of simple outcome measures such as those obtained from kinematic characteristics (see Section \ref{sec:TaskStats} and \ref{sec:KinematicClassification}) and on the other hand demonstrates the discriminative power of dynamical characteristics obtained here using a PWARX model. The latter provides a more detailed segmentation and insights into the dynamic make-up of the behavior and their spatial organization. This more detailed information provides correlation with important procedural movement stages, yet it requires no prior, high-level knowledge about the task to be implemented. Finally, the spatial characteristics of the segmented performance data provides a measure of the ability to organize the different stages of behavior in a manner which is consistent with the spatial and dynamic constraints of the task and operator skills. These results demonstrate that dynamical segmentation techniques can access attributes across the entire process hierarchy and provide the foundation to a more comprehensive skill analysis and modeling framework. Future work will immediately extend this approach to more FLS tasks and ultimately to different tasks in laparoscopic, robotic, and open surgery and incorporate gaze characteristics.

\bibliographystyle{IEEEtran}
\bibliography{M2CAI_Surgery_Skill_Analysis}

\end{document}